\documentstyle[sprocl]{article}

\def\be{\begin{equation}}
\def\ee{\end{equation}}
\def\bea{\begin{eqnarray}}
\def\eea{\end{eqnarray}}
\def\f{\frac}
\def\p{\partial}
\def\ni{\noindent}
\def\l[{\left[}
\def\r]{\right]}
\begin{document}

\title{VACUUM RADIATION IN CONFORMALLY INVARIANT QUANTUM FIELD THEORY}

\author{ V. ALDAYA, M. CALIXTO}
\address{University of Granada/CSIC, 18002 Granada, Spain.}
\author{ J.M. CERVER\'O }
\address{University of  
Salamanca, 37008 Salamanca, Spain.}


\maketitle\abstracts{Although the whole conformal group 
$SO(4,2)$ can be considered as a symmetry
in a classical massless field theory, the subgroup of special conformal 
transformations (SCT), usually related to transitions to uniformly 
accelerated frames, causes vacuum radiation in the corresponding quantum 
field theory, in analogy to the Fulling-Unruh
effect. The spectrum of the outgoing
particles can be calculated exactly and proves to be a generalization 
of the Planckian one. }

The conformal group $SO(4,2)$ has ever been recognized as a symmetry of the 
Maxwell equations for {\it classical} electro-dynamics 
\cite{Maxwell}, and more recently considered as 
an invariance of general, non-abelian, maseless gauge field theories at 
the classical 
level. However, the {\it quantum} theory raises, in general, 
serious problems in 
the implementation of conformal symmetry, and much work has been devoted to 
the study of the physical reasons for that (see e.g. Ref. \cite{Fronsdal}). 
Basically, the main trouble 
associated with this quantum symmetry (at the second quantization level) 
lies in the difficulty of finding 
a vacuum,  which is {\it stable} under special conformal 
transformations acting on 
the Minkowski space in the form:
\be
x^\mu \rightarrow {x'}^\mu=\f{x^\mu+c^\mu
x^2}{\sigma(x,c)},\;\;\;\sigma(x,c)=1+2c x+c^2 x^2
\label{confact}.
\ee
\ni These transformations, which can be interpreted as transitions to systems 
of relativistic, uniformly accelerated observers (see e.g. Ref. \cite{Hill}), 
cause {\it vacuum radiation}, a phenomenon 
analogous to the Fulling-Unruh effect
\cite{Fulling,Unruh} in a non-inertial reference  frame. To be more precise, 
if $a(k),a^+(k)$ are the Fourier components of a scalar massless field 
$\phi(x)$, satisfying the equation 
\be 
\eta^{\mu\nu}\p_{\mu}\p_\nu \phi(x)=0\,,
\ee
\ni  then the Fourier components 
$a'(k), {a'}^+(k)$ of the transformed field 
$\phi'(x')=\sigma^{-l}(x,c)\phi(x)$ by (\ref{confact}) ($l$ being the
conformal dimension) are expressed in terms of both $a(k),a^+(k)$ through a 
Bogolyubov transformation
\be
a'(\lambda)=\int{dk\l[ A_\lambda(k)a(k)+B_\lambda(k)a^+(k)\r]}\,. \label{bog}
\ee
\ni In the second quantized theory,  the vacuum states defined by the
conditions $a(k)|0\rangle   =0$ and $a'(\lambda)|0'\rangle   =0$, 
are not identical if the coefficients $B_\lambda(k)$ in (\ref{bog})
are not zero. In this case the new vacuum has a non-trivial content of 
untransformed particle states.
 
This situation is always present when quantizing field theories in
curved space as well as in flat space, whenever some kind of global mutilation
of the space is involved. This is the case of the 
natural quantization in Rindler
coordinates \cite{Fulling}, which leads to a quantization inequivalent to the
normal Minkowski quantization, or that of a quantum field in a box, where a
dilatation produces a rearrangement of the vacuum \cite{Fulling}. 
Nevertheless,
it must be stressed that the situation for SCT is more peculiar. 
The rearrangement of the vacuum in a massless
QFT due to SCT, even though they are a 
symmetry of the classical system, behaves as if the conformal group were 
{\it spontaneously broken}, and this fact can be 
interpreted as a kind of topological 
{\it anomaly}.

Thinking of the underlying reasons for this anomaly, we are tempted to make
the singular action of the transformations (\ref{confact}) in Minkowski space
responsible for it, as has been in fact pointed out in \cite{rusos}. 
However, a
deeper analysis of the interconnection between symmetry and quantization  
reveals a more profound obstruction to the 
possibility of implementing unitarily 
STC  in a  generalized Minkowski space (homogeneous space of the 
conformal group), free 
from singularities. 
This obstruction is traced back to the 
impossibility of representing the entire 
$SO(4,2)$ group unitarily and irreducibly on a space of functions depending 
arbitrarily on $\vec{x}$ (see e.g.  Ref. \cite{Fronsdal}), so that a Cauchy 
surface determines the evolution in 
time. Natural representations, however, can be constructed by means of wave 
functions having support on the hole space-time and evolving in some kind of 
{\it proper time}. 
It is proved (see \cite{conforme}) that 
unitary irreducible representations of the conformal 
group require the generator $P_0$ of time translations to have dynamical 
character (i.e., it has a canonically conjugated pair), as it happens with 
the spatial component $P_1$, due to the appearance 
of a central term $\hat{1}$ in the quantum conmutators  
\be
[P_\mu,K_\nu]= 
-\eta_{\mu\nu}(2D+4N\hat{1})
\ee
\ni  ($K_\nu$ and $D$ denote the generators of SCT 
and dilatations, respectivelly) proportional to a parameter $N$, which 
characterizes the unitary irreducible 
representations of the conformal group. So, 
conformal wave functions 
$\psi^{(N)}$ have 
support on the whole space-time. If we  forced the funtions $\psi^{(N)}$ to 
evolve in time according to the Klein-Gordon-like equation 
\be
Q\psi^{(N)}=P_\mu P^\mu \psi^{(N)}=0\,,\label{nullmass}
\ee
\ni  (``null square mass condition'', i.e., by selecting those 
functions nullified by the 
Casimir operator $Q$ of the Poincar\'e subgroup of $SO(4,2)$) we would find 
that the appearance of {\it quantum} terms proportional to $N$ at 
the right-hand 
side of 
the quantum commutators 
\be
\l[K_\mu,Q\r]= f_\mu(x,t)Q +8NP_\mu
\ee
\ni (where $f_\mu(x,t)$ are 
 some funtions on the generalized Minkowski space),
terms  which do not appear at the classical level 
($N=0$), prevent the whole conformal group to be an exact symmetry of the 
massless quantum field. This way, the quantum time evolution  
itself destroys the conformal symmetry, 
leading to some sort of {\it dynamical symmetry breaking} which 
preserves the Weyl 
subgroup (Poincar\'e + dilatations). The SCT do not leave the 
Eq. (\ref{nullmass}) invariant, and this 
fact  manifests, at the second quantization level, through a {\it radiation} 
of the vacuum of the massless quantum field (``Weyl vacuum'') under the 
action of SCT, i.e., from the point of view of an uniformly accelerated 
observer. The spectrum of the outgoing particles can be calculated exactly 
\cite{conforme} and proves to be a generalization of the Plankian one, this 
recovered in the limit $N\rightarrow 0$. The temperature of this thermal 
bath is linear in the acceleration parameter, as in the reference 
\cite{Unruh}.

\section*{Acknowledgments}
Work partially supported by the DGICYT under contracts PB92-1055,
PB92-0302, PB95-1201 and PB95-0947

\end{document}